\documentclass[letterpaper, 10 pt, conference]{ieeeconf}  
\IEEEoverridecommandlockouts                 
\usepackage{graphics} 
\usepackage{graphicx}
\usepackage{epsfig} 
\usepackage{mathptmx} 
\usepackage{times} 
\usepackage{amsmath} 
\usepackage{amssymb}  
\usepackage{hyperref}
\usepackage{tabularx}
\usepackage{adjustbox}
\usepackage{color}
\renewcommand{\thetable}{\arabic{table}}

\title{\LARGE \bf
NBA2Vec: Dense Feature Representations of NBA Players
}

\author{Webster Guan$^{1}$, Nauman Javed$^{2}$, and Peter Lu$^{3}$%
\thanks{$^{1}$Department of Chemical Engineering, Massachusetts Institute of Technology, Cambridge, MA, USA. 
        {\tt\small \href{mailto:wjguan@mit.edu}{wjguan@mit.edu}}}%
\thanks{$^{2}$Harvard-MIT Health Sciences and Technology, Cambridge, MA, USA.
        {\tt\small \href{mailto:njaved@mit.edu}{njaved@mit.edu}}}%
\thanks{$^{3}$Department of Physics, Massachusetts Institute of Technology, Cambridge, MA, USA.  
        {\tt\small \href{mailto:lup@mit.edu}{lup@mit.edu}}}%
}

\begin{document}

\maketitle
\thispagestyle{plain}
\pagestyle{plain}

\begin{abstract}

Understanding a player's performance in a basketball game requires an evaluation of the player in the context of their teammates and the opposing lineup. Here, we present \textit{NBA2Vec}, a neural network model based on \textit{Word2Vec} \cite{word2vec} which extracts dense feature representations of each player by predicting play outcomes without the use of hand-crafted heuristics or aggregate statistical measures. Specifically, our model aimed to predict the outcome of a possession given both the offensive and defensive players on the court. By training on over 3.5 million plays involving 1551 distinct players, our model was able to achieve a 0.3 K-L divergence with respect to the empirical play-by-play distribution. The resulting embedding space is consistent with general classifications of player position and style, and the embedding dimensions correlated at a significant level with traditional box score metrics. Finally, we demonstrate that NBA2Vec accurately predicts the outcomes to various 2017 NBA Playoffs series, and shows potential in determining optimal lineup match-ups. Future applications of NBA2Vec embeddings to characterize players' style may revolutionize predictive models for player acquisition and coaching decisions that maximize team success.

\end{abstract}

\section{INTRODUCTION}

Successful coaches construct optimal lineups for given situations in basketball games based on a deep understanding of each player's play-style, strengths, and weaknesses in the context of all other players on the court. Studying the distribution of contexts and their outcomes in which a player takes part may provide insights into aspects of player's performance and play style that are not otherwise reflected in traditional basketball statistics. While much of basketball analytics relies on the use of hand-crafted advanced statistics (e.g. Wins Above Replacement and Offensive/Defensive rating) and aggregate statistics (e.g. FG\%, assists), they tend to not capture these contextual influences and effects not present in box scores. Models capable of characterizing players based on these contextual factors would offer greater insight into individual player performance and play-style, and may shed light on how to construct optimal lineups for specific situations. Constructing such frameworks may be possible given the wealth of play-by-play game data and recent advances in machine learning and natural language processing (NLP) algorithms. 

In particular, the problem of generating accurate representations of players in sports analytics is analogous to the problem of word embeddings in NLP. Word embedding models aim to create real-valued vector embeddings of words that encode complex semantic structure. {\em Word2Vec} is a class of word embedding models that extract useful features of words given the sentences, known as the ``context," in which the words are commonly used \cite{word2vec}. This allows Word2Vec to be trained in an unsupervised way on a large corpus of written works and extract meaningful relationships. Once trained, the word embeddings can then be applied to a variety of other tasks as a pretrained initial transformation in manner analogous to transfer learning.

\begin{figure*}[htp]
\centering
\centerline{\includegraphics[width=\linewidth]{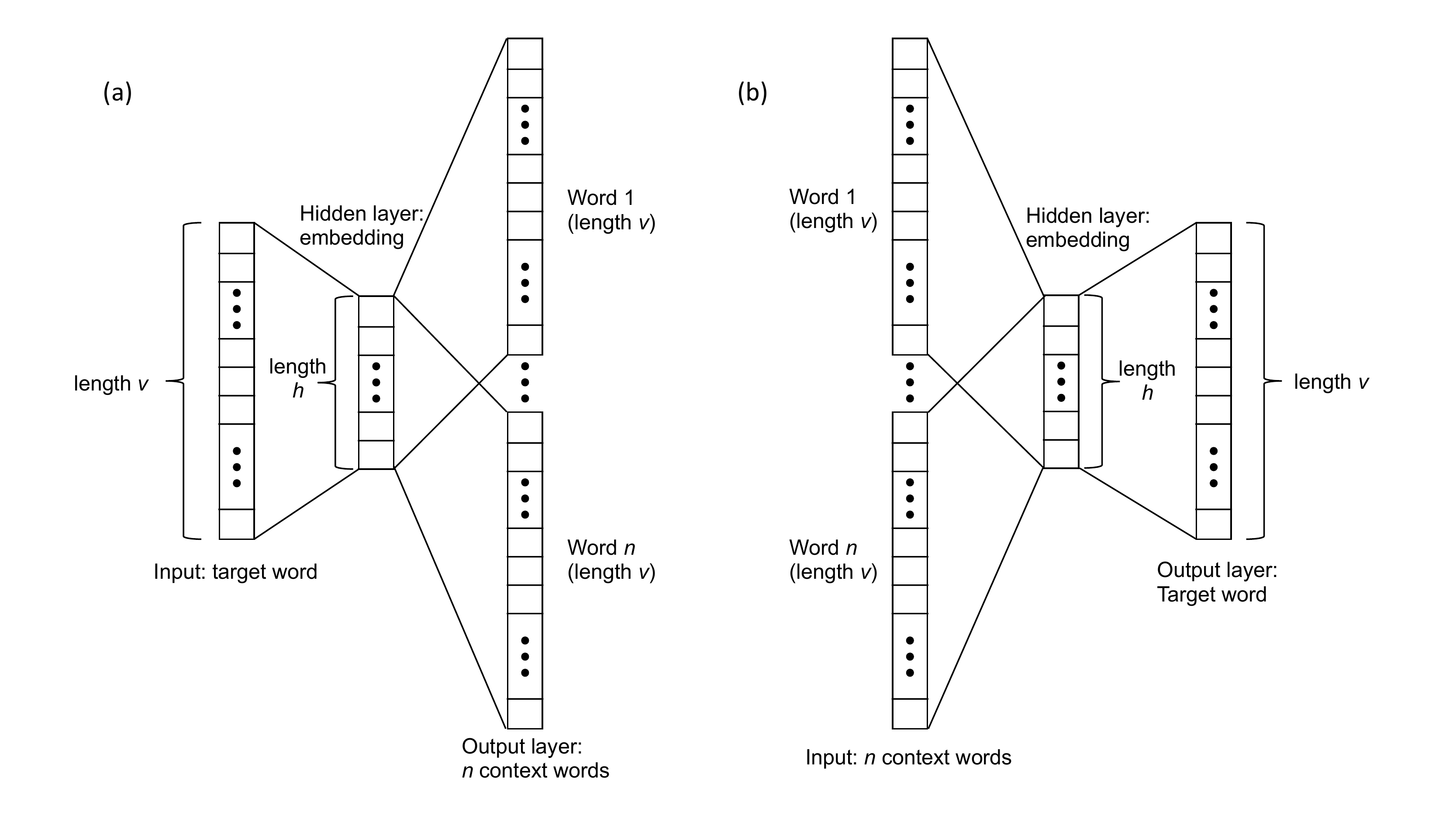}}
\caption{(a) Skip-gram word embedding model. The embedding is extracted from the hidden layer after training to a target of \textit{n} context words. \textit{v} represents the vocabulary size, or length of the word vectors, while \textit{h} represents the embedding length. (b) CBOW model. A given word's embedding is computed by averaging the hidden layer representations of all contexts in which it appears.}
\label{fig:1.1}
\end{figure*}

In the training phase, the way in which the context of each word is used can be different; in particular, Word2Vec uses either a continuous bag-of-words (CBOW) model, or a skip-gram model. The skip-gram method (Figure \ref{fig:1.1}a) takes the target word as input to a neural network's hidden layer(s) and attempts to predict the words that immediately surround the target word in a given sentence. On the other hand, the CBOW method (Figure \ref{fig:1.1}b) takes the words around a target word as input, predicting the target word as output. The result of training these models is a dense vectorial word representation that captures the meaning of each word. For example, Word2Vec finds more similarity between the words ``king" and ``queen" than between ``king" and ``vindication." The ability of word embeddings to accurately capture the relationship and analogies among words is shown by Word2Vec arithmetic: for instance, \textit{Paris $-$ France $+$ Italy $=$ Rome}. 

The success of word embeddings in NLP has inspired its recent application in sports analytics to characterizing batter and pitcher performance and play style in baseball \cite{mlb2vec}. In that study, a neural network was trained using pitcher-batter matchups as inputs and the outcome of each at-bat (e.g. single, home run, flyout to right field) as outputs to create the player embeddings. The author was able to successfully visualize macroscopic differences in embedding clusters (e.g. right-handed vs. left-handed vs. switch hitters, power hitters vs. on-base hitters) and model previously unseen at-bat matchups, suggesting that the word embedding concept may be feasible and promising for creating player representations.  
 
In this study, we applied this concept to extract representations of different NBA players by producing a embeddings of every player, which we term {\em NBA2Vec}. Similar to \cite{mlb2vec}, the embedding for each player was generated by training a neural network aimed at predicting the outcome of each possession given the ten players on the court. Unlike in \cite{mlb2vec}, which takes advantage of the mostly one-on-one nature of baseball dynamics, we used all players on the court to ensure accurate modeling of holistic relationships and dynamics between players on the same and different teams. This increased the complexity of the network, and due to the n-body nature of the problem, required the network to exhibit permutation invariance. Unlike previous attempts to generate NBA player representations \cite{lutz2012cluster} using purely high level, aggregate statistics and hand-picked features (e.g. number of shots taken in the paint, FG\%, FT\%, assists, etc.), our embedding approach learns directly from raw play-by-play data by automatically generating rich features which account for the ``context" that affects a player’s style and statistics. The latent features encoded in these player embeddings can shed light on both the play style and effectiveness of different types of players, and can be used as inputs for further downstream processing and prediction tasks. 

\section{METHODS}

\subsection{Data Sets and Preprocessing}

We used play-by-play and players-on-court data provided by the NBA, which featured over 9 million distinct plays, with 1551 distinct players taking the court in these plays. To create the input to the network, we denoted each player with an index from 0 to 1550. For the outputs to the network, we needed to encode the possible outcomes of each play. In order to encourage learning, we only considered key outcomes, omitting rebounds and defensive plays. We chose to use 23 distinct outcomes, some examples of which are included in Table \ref{tab:2.1}. The provided raw play-by-play data was more specific on outcomes of plays, but we grouped many of these plays (e.g. ``driving layup shot, dunk shot, reverse dunk shot, hook shot" were all considered ``close-up shots") together for simplicity. This preprocessing resulted in 4.5 million plays, of which we used 3.7 million for a training set and the remainder as a validation set. We used the Pandas library to preprocess the data \cite{mckinney2010data}.

\begin{table} [htp!]
    \centering
    \caption{Types of play outcomes.}
    \begin{tabular}{|c | c |}
        \hline
         Play Index & Play Outcome\\
         \hline
        0 & Mid-range jump shot made\\ 
        1 & Mid-range jump shot missed\\
        2 & Mid-range jump shot made + 1 free throw made \\
        3 & Mid-range jump shot made + 1 free throw missed \\
        \hline
        4 & Close-range shot made \\
        5 & Close-range shot missed\\
        6 & Close-range shot made + 1 free throw made\\
        7 & Close-range shot made + 1 free throw missed\\
        \hline 
        8 & 0/1 FT made \\
        9 & 1/1 FT made \\
        10 & 0/2 FT made \\
        11 & 1/2 FT made \\
        12 & 2/2 FT made \\
        13 & 0/3 FT made \\
        14 & 1/3 FT made \\
        15 & 2/3 FT made \\ 
        16 & 3/3 FT made \\
        \hline
        17 & 3PT shot made \\
        18 & 3PT shot missed \\
        19 & 3PT shot made + 1 free throw made \\
        20 & 3PT shot made + 1 free throw missed \\
        \hline 
        21 & Turnover  \\ 
        22 & Foul \\
        \hline  
    \end{tabular}
    \label{tab:2.1}
\end{table}

\subsection{NBA2Vec: Player Embedding Model}

\begin{figure*}[htp]
\centering
\centerline{\includegraphics[width=\linewidth, trim=0 0 130 0, clip]{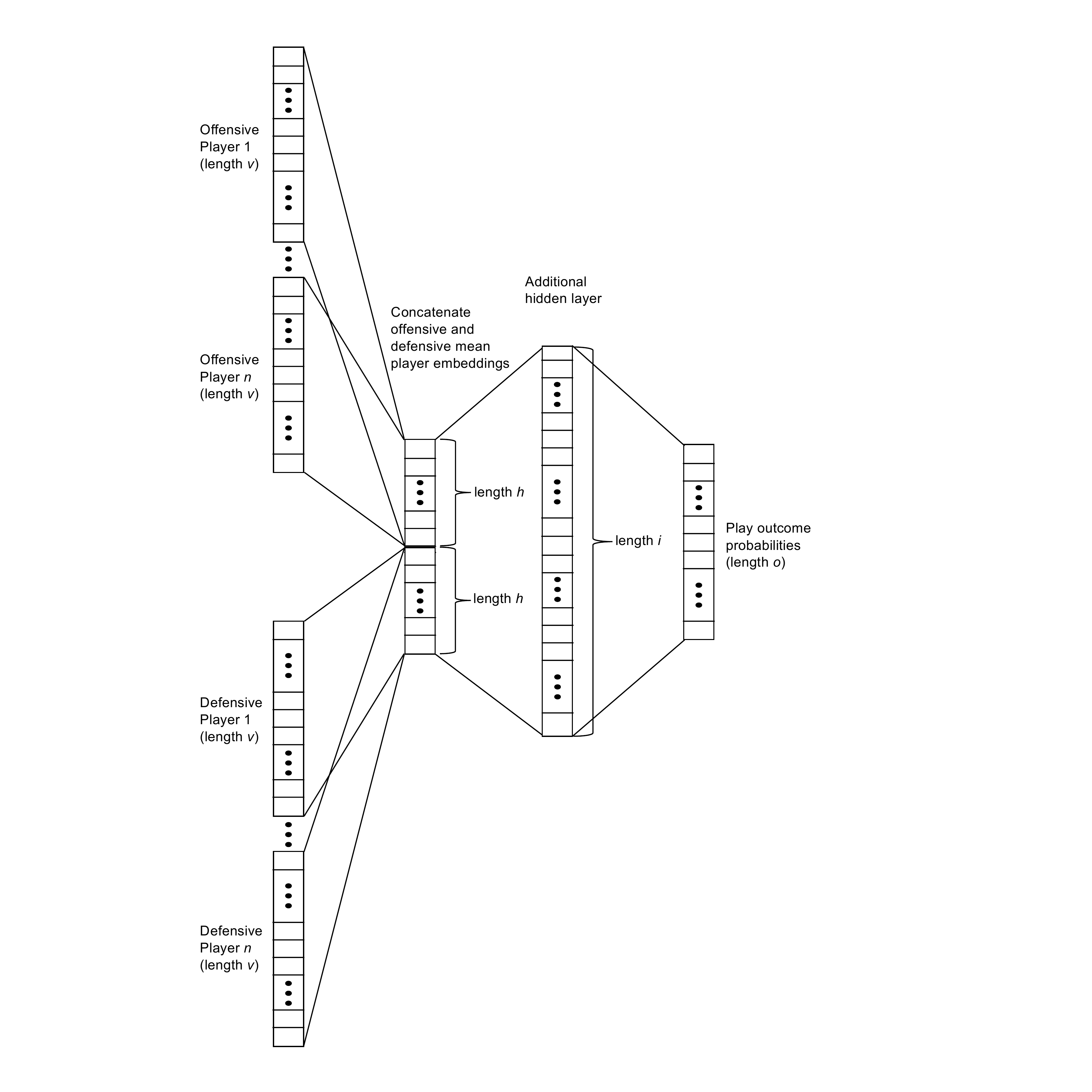}}
\caption{NBA2Vec model architecture. We have $v= 1551$ players, which are mapped to $h=8$ dimensional embeddings. After the $n=5$ offensive and defensive player embeddings are separately averaged and then concatenated, they are fed through an $i=128$ hidden layer with a ReLU activation. The final output layer with a softmax activation predicts a probability distribution over $o=23$ play outcomes.}
\label{fig:2.1}
\end{figure*}

To train informative embeddings for players in the NBA, we created a neural network architecture that predicts the distribution of play outcomes given a particular offensive and defensive lineup (Figure \ref{fig:2.1}). For each play, we first embed the 10 players on the court using an 8 dimensional shared player embedding. We then separately average the 5 offensive and 5 defensive player embedding vectors. These two mean player embeddings (i.e. an offensive and a defensive lineup embedding) are concatenated and fed through one additional hidden layer of size 128 with a ReLU activation before outputting 23 outcome scores. Applying a softmax activation to the scores produces probabilities that we interpret as the distribution of play outcomes (Table \ref{tab:2.1}). The entire network is trained end-to-end with a cross entropy loss function that stochastically minimizes the K--L divergence between the true play outcomes from the data and the predicted distribution from the model. This model was built and trained using the PyTorch framework \cite{paszke2017automatic}.

\subsection{Validation and Post-processing}

\subsubsection{Validation of NBA2Vec}
To evaluate the efficacy of the NBA2Vec model used to generate the embeddings, we characterized the difference between the predicted and empirical distributions of play outcomes. The Kullback--Leibler (K--L) divergence was used as the metric to compare the distributions. K--L divergence is given by 
\begin{equation}
D_{KL}\left(p\middle\|q\right) = \sum_{i=1}^{N}{p(x_{i})\log{\frac{p(x_{i})}{q(x_{i})}}}.
\label{eq:3.1}
\end{equation}
This measures the number of encoded bits lost when modeling a target distribution $p(x)$ (in this case, the empirical distribution) with some approximate distribution $q(x)$ (in this case, our predictive model). Thus, a low K--L divergence value ($D_{KL} \approx 0$) implies a good approximation of the target distribution, while a larger value ($D_{KL} \gg 0$) implies poor approximation. 

Due to the large number of unique lineup-matchup combinations, some of which do not appear enough for a proper empirical distribution to be generated, we decided to only look at K--L divergences for lineup-matchup combinations with more than 15 plays. This analysis was performed on the last 25 games of the data set (corresponding to the last 25 playoff games in the 2018 NBA playoffs, and 5102 plays).

\subsubsection{Embedding Analysis\label{sec:embed}}
After training the model, we extracted the shared embedding layer and use dimensionality reduction, clustering, and visualization methods to explore the learned player embeddings. In particular, we used t-SNE---a dimensionality reduction method based on local neighbor structure---to visualize our 8 dimensional embeddings in 2 dimensions \cite{maaten2008visualizing}. We also used 2D principal component analysis (PCA) for dimensionality reduction before performing k-means clustering. PCA is a statistical method that uses orthogonal decomposition to transform a set of observations of correlated variables into a set of observations of uncorrelated variables, or principal components. Each successive principal component explains the maximal variance in the data while remaining orthogonal to the preceding component. K-means clustering is a simple clustering method that aims to partition n observations into k clusters, where each observation belongs to the cluster with the nearest mean. Our approaches are further described in section \ref{embedding_analysis}. These dimensionality and clustering techniques were conducted using implementations from the Scikit-learn library \cite{scikit-learn}.

\subsubsection{Exploring Lineup Combinations}
 
To further explore the macroscopic predictive nature allowed by these embeddings, we used the neural network model to predict the outcomes of games based on each team's most frequent 5-player lineup. For each pair of teams, the model would output the distribution of possible play outcomes. We would then sample these distributions to determine which plays would occur in a given game, and based on this, predict a game score. Assuming 100 possessions for each team and that no substitutions are ever made, we ran the model on various playoff series match-ups from the 2016-17 season, simulating 1000 best-of-7 series between each pair of teams. Certain playoff series were not simulated because the most frequent lineups contained players that were not among the raw data's most common 500 players. 

\section{RESULTS AND DISCUSSION}

\subsection{Embedding Analysis\label{embedding_analysis}}

In order to better understand the generated player representations, we used t-SNE (described in section \ref{sec:embed}) to visualize the 8 dimensional feature vector in 2D. As depicted in the 2D t-SNE plot (described in section \ref{sec:embed}) {\ref{fig:3}, we see that the centers and guards separate nicely, while forwards (yellow) are scattered throughout both groups. This is consistent with our intuition about the roles and play-styles of these classes of players---guards and centers fulfill very distinct roles while forwards can be more center-like or guard-like and fulfill multiple roles (i.e. LeBron James). Despite t-SNE's utility in preserving high dimensional local structure, it is not effective at preserving the global structure of data, which along with the high dimensionality of our player representations, may help explain why it is difficult to visualize clear separation between different groups of players. 

To further characterize the learned embeddings, we k-means clustered 8 scaled and centered embedded dimensions. We decided to select 3 clusters for our initial analysis after observing a decreased rate of decline in the variance as a function of the number of clusters, as seen in the elbow plot in Figure \ref{fig:4}. In Figure \ref{fig:5} we see that k-means yields 3 distinct clusters, which roughly group players with similar roles/playing styles. For example, the yellow cluster seems to correspond to guards, the green cluster with centers/forwards, and the red cluster with forwards. Note that because of the high dimensional nature of our player representations and the fact that we are projecting players onto two dimensions, the euclidean distance between points on the shown plot is not entirely representative of player similarity. 

The observed clusters also seem to suggest that as opposed to fitting neatly within the traditional 5 positions of basketball, players actually perform diverse roles that would place them in multiple categories. As seen in figure \ref{fig:5}, each of the clusters comprises multiple groups of players---clusters 1 and 2 comprise positions 1-5, while cluster 2 corresponds mostly to centers/forwards. Roughly, this may reflect that successful players are rarely one trick ponies---centers must be able to shoot, and point guards must be able to score. In general, we can observe that the learned embeddings roughly correspond to general basketball intuition. However, the embeddings are also capturing player characteristics that may not be entirely reflected in traditional metrics such as box score.

Exploring the structure of the embedded space by calculating the nearest neighbors by distance for various players further validates the learned player representations. For example, Chris Paul, a canonical point guard, has nearest neighbors including other point guards such as Steve Nash, Jose Calderon, and Jason Terry. Shaquille O'Neal, a classic big man, has nearest neighbors including centers such as Dwight Howard, Roy Hibbert, Tiago Splitter, and Rudy Gobert. 

Next, we calculated Pearson's correlation between the top two PCA dimensions and player metrics including minute adjusted rates for field goals made, three pointers, assists, rebounds, and plus---minus, as shown in figure \ref{fig:7}. Rudimentary analysis revealed that PCA dimension 1 correlated at a significant level (corrected $\alpha = 5\times 10^{-4}$) with rebounds, assists, and three pointers, while PCA dimension 2 correlated at a significant level with rebounds and assists (corrected $\alpha = 5\times 10^{-4}$). For both dimensions, the noted correlations remained significant even with Bonferroni adjustment.

Our exploratory analysis of the embeddings reveals that the learned player representations encode meaningful information that correspond roughly to our intuition about various players. Through a rudimentary analysis, the embedded dimensions seem to correspond to a complex combination of player characteristics as well as real player performance metrics. 

\begin{figure}[htb!]
    \centering
    \includegraphics[width=\linewidth, trim=10 0 10 0, clip]{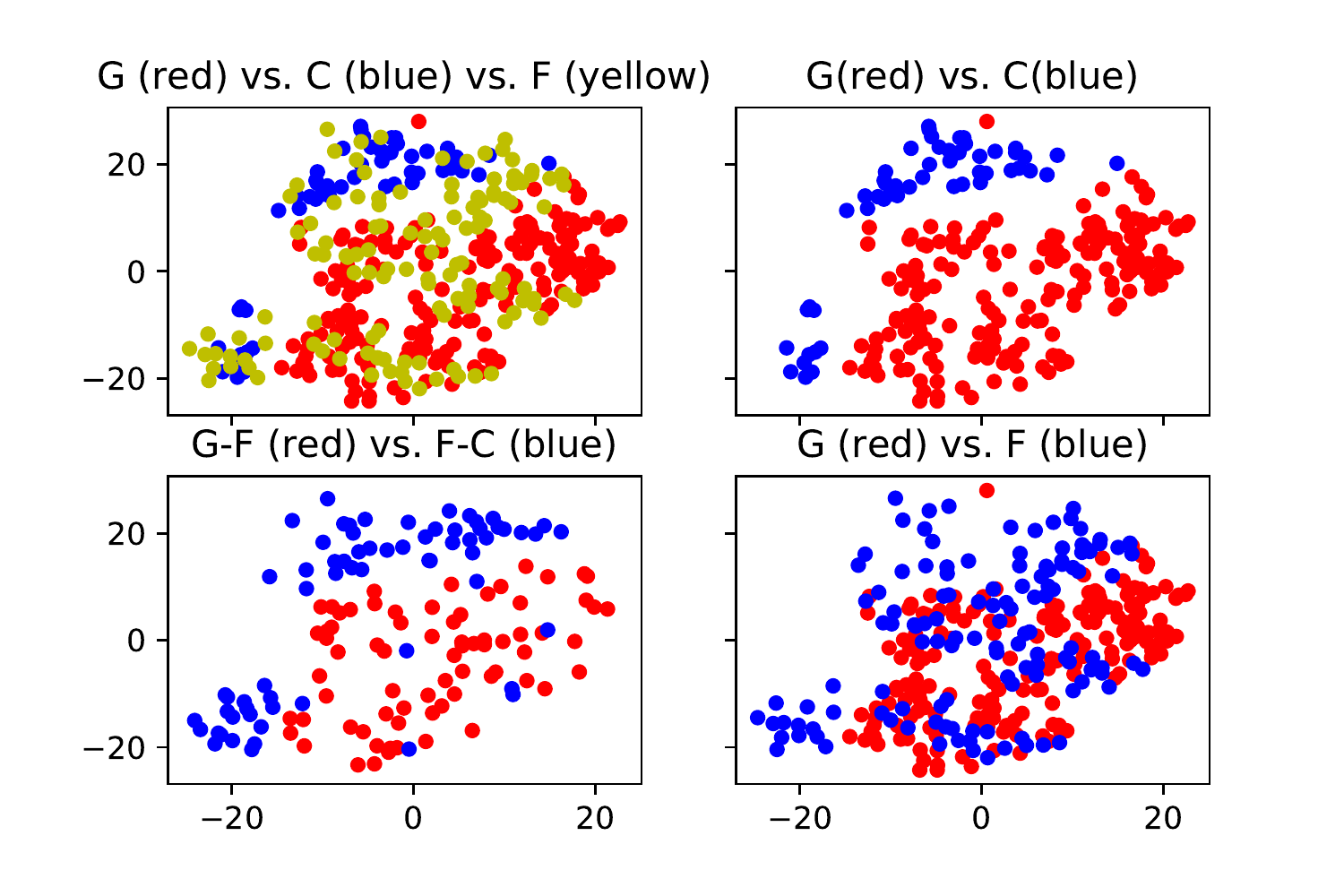}
    \caption{Two-dimensional t-SNE visualization of player representations, colored by position. (G = guards, C = centers, F = forwards, G-F = guard-forwards, F-C = forward-centers).}
    \label{fig:3}
\end{figure}

\begin{figure}[htb!]
    \centering
    \includegraphics[width=\linewidth]{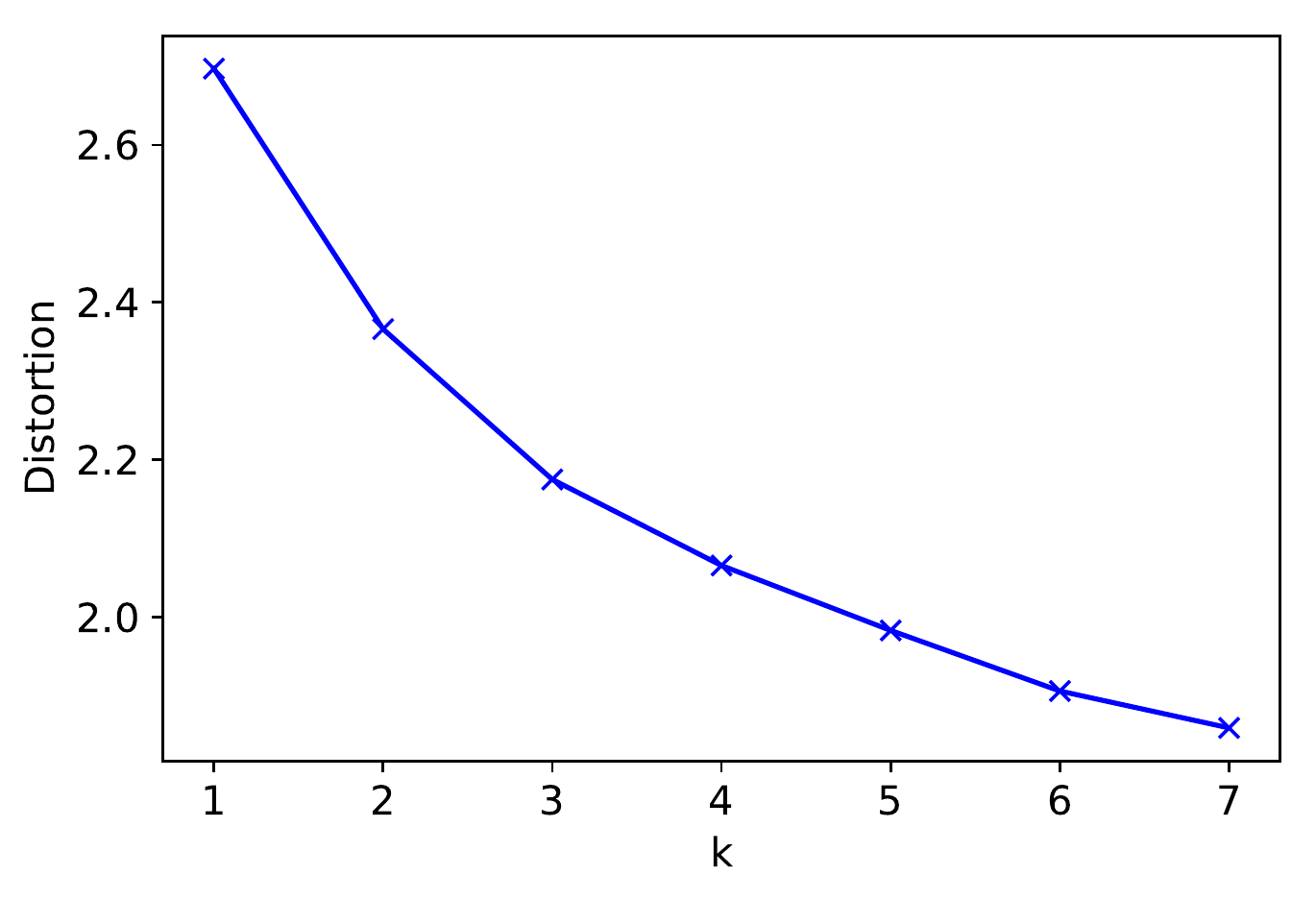}
    \caption{Elbow method showing variance in data as a function of the number of clusters in order to optimize $k$.}
    \label{fig:4}
\end{figure}

\begin{figure}[htb!]
    \centering
    \includegraphics[width=\linewidth]{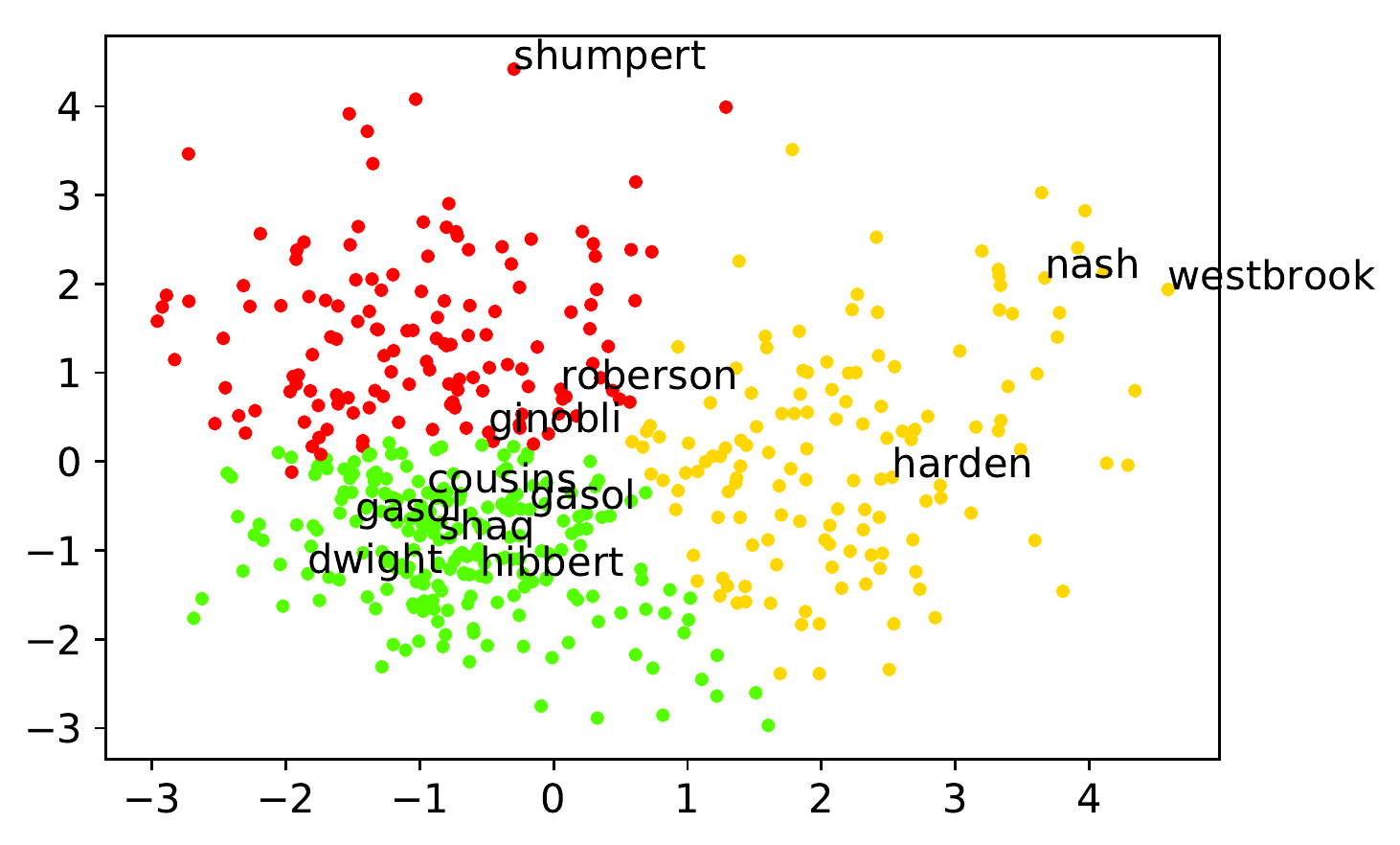}
    \caption{Points projected on to first two principal components, colored by clusters identified by k-means clustering with $k=3$.}
    \label{fig:5}
\end{figure}

\begin{figure*}[htb!]
    \centering
    \includegraphics[width=0.79\linewidth]{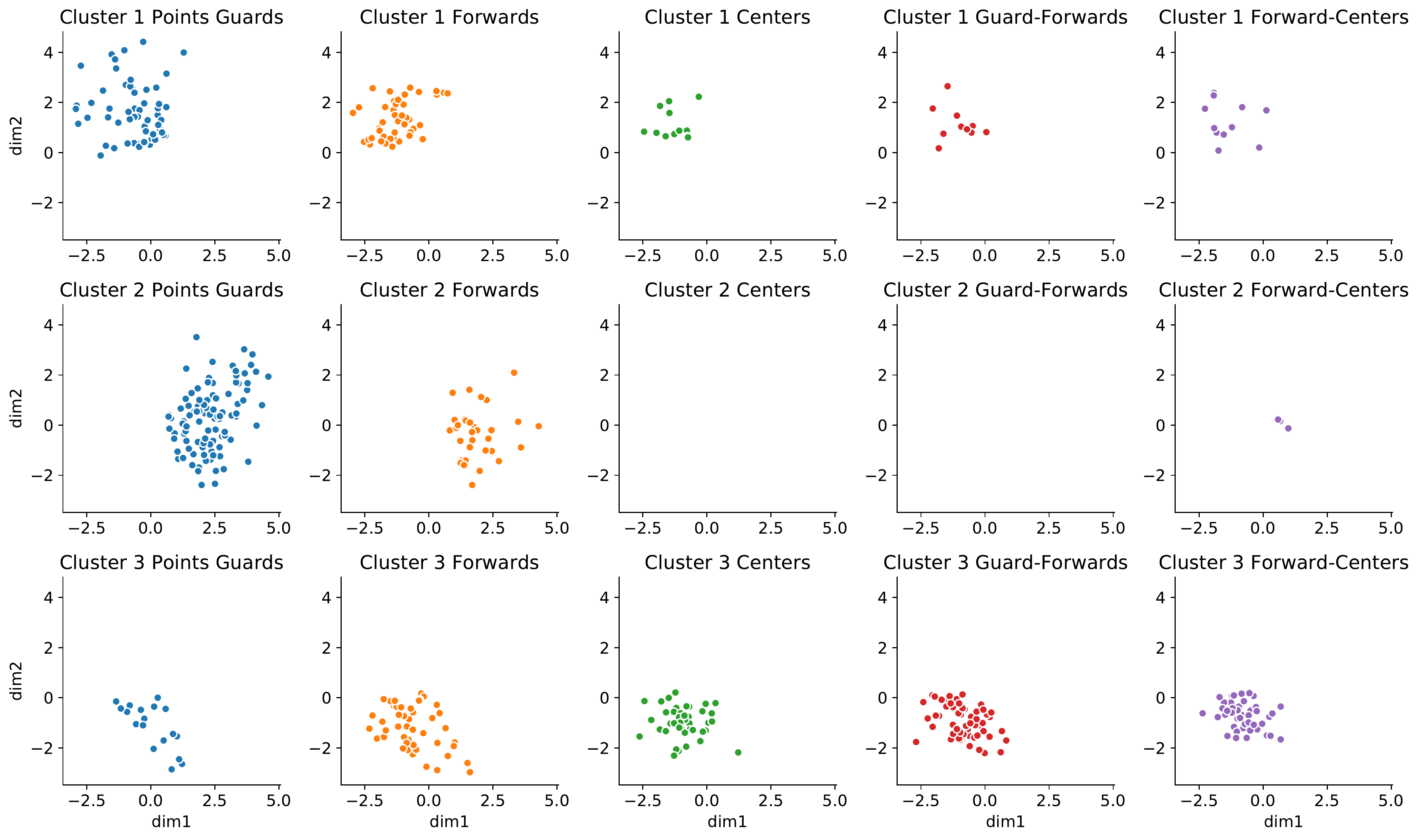}
    \caption{Different types of players comprising each identified cluster in Figure \ref{fig:5}. }
    \label{fig:6}
\end{figure*}

\begin{figure}[htb!]
    \centering
    \includegraphics[width=\linewidth]{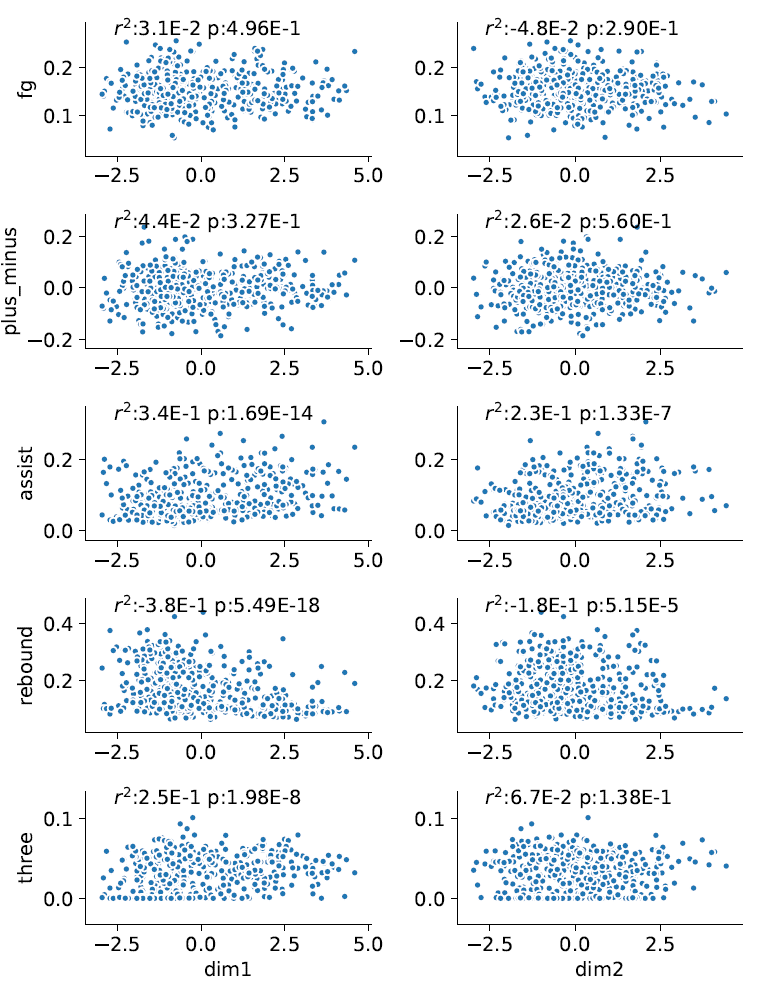}
    \caption{Correlations and p-values of different metrics with the two top PCA dimensions for each player. Each raw metric is summed for each player and normalized by the total number of minutes played. dim1 = PCA dimension 1, dim2 = PCA dimension 2, fg = field goal. Reported p-values are not Bonferroni corrected but, after correction, remain significant at $\alpha =5\times 10^{-4}$ with 5 comparisons.}
    \label{fig:7}
\end{figure}

\subsection{Validation of NBA2Vec}

Validation of the NBA2Vec network was performed on plays in the final 25 games of the data set, and a mean K--L divergence of $0.301 \pm 0.162$ was achieved. Some example predicted vs. empirical distributions of play outcomes are shown in Figure \ref{fig:3.1}, showing that the model is able to closely approximate the target distribution. 

We also wanted to determine the minimum number of plays needed to create an accurate empirical distribution that can be modeled by the predictive network. Plotting the K--L divergence vs. number of plays used in the empirical distribution (Figure \ref{fig:3.2}), we can estimate the minimum number of plays to reach the minimum K--L divergence to be around 30.  

\begin{figure}[htb!]
    \centering
    \includegraphics[scale=0.625]{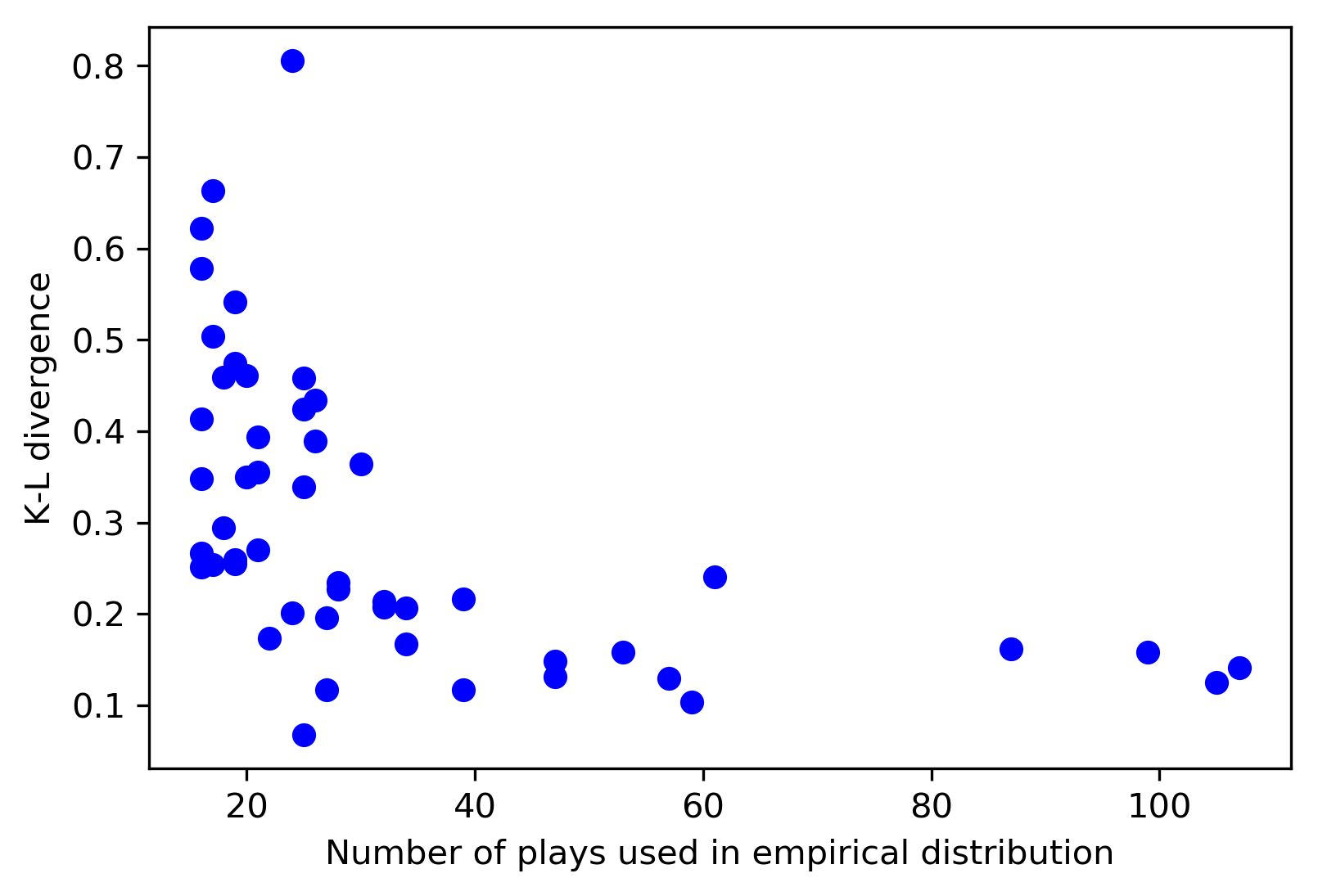}
    \caption{K--L divergence vs. Number of plays used in empirical distribution. The K--L divergence reaches a minimum plateau after about 30 plays. }
    \label{fig:3.2}
\end{figure}

\begin{figure*}[!htp]
    \centering
    \includegraphics[scale=0.425]{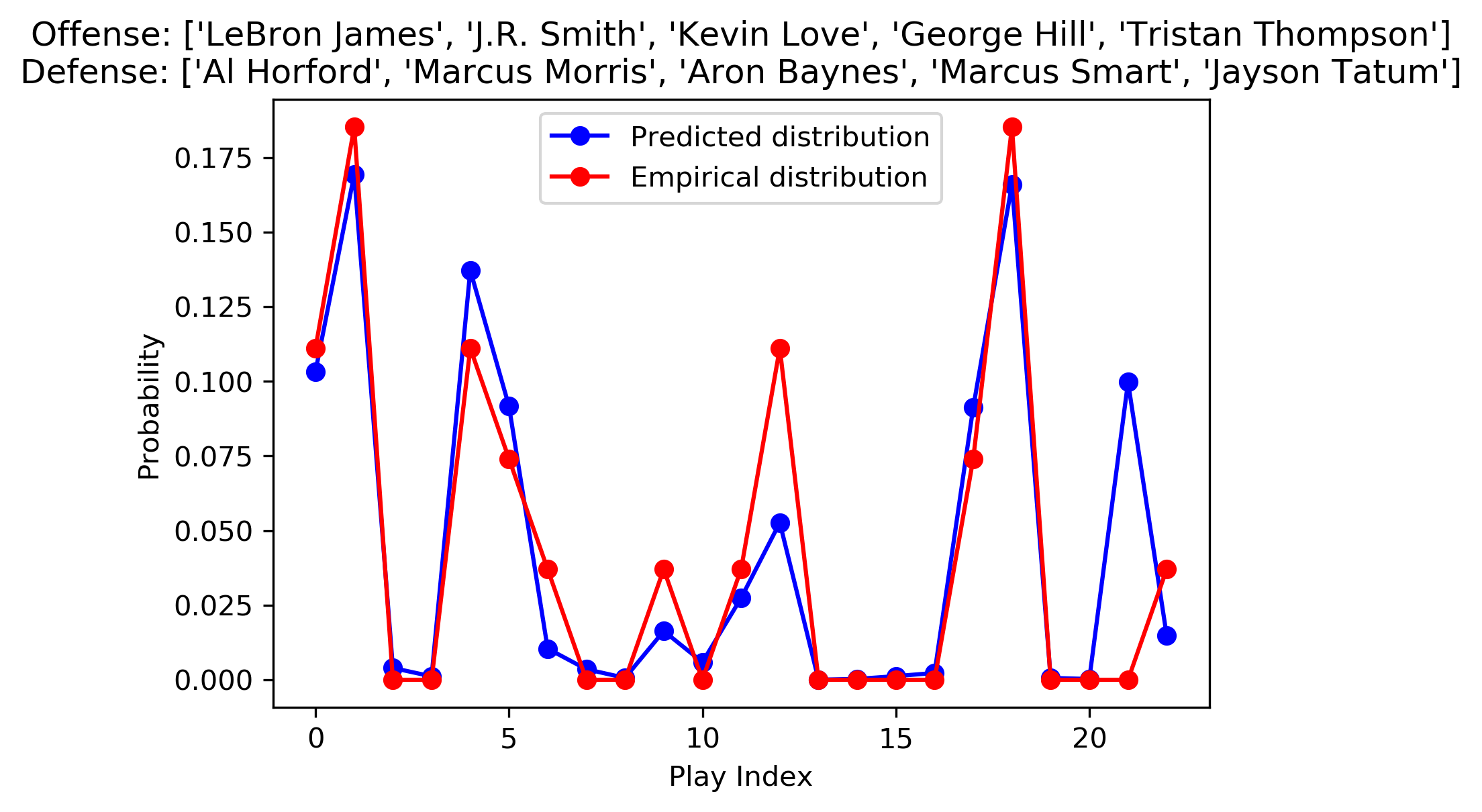}
    \includegraphics[scale=0.425]{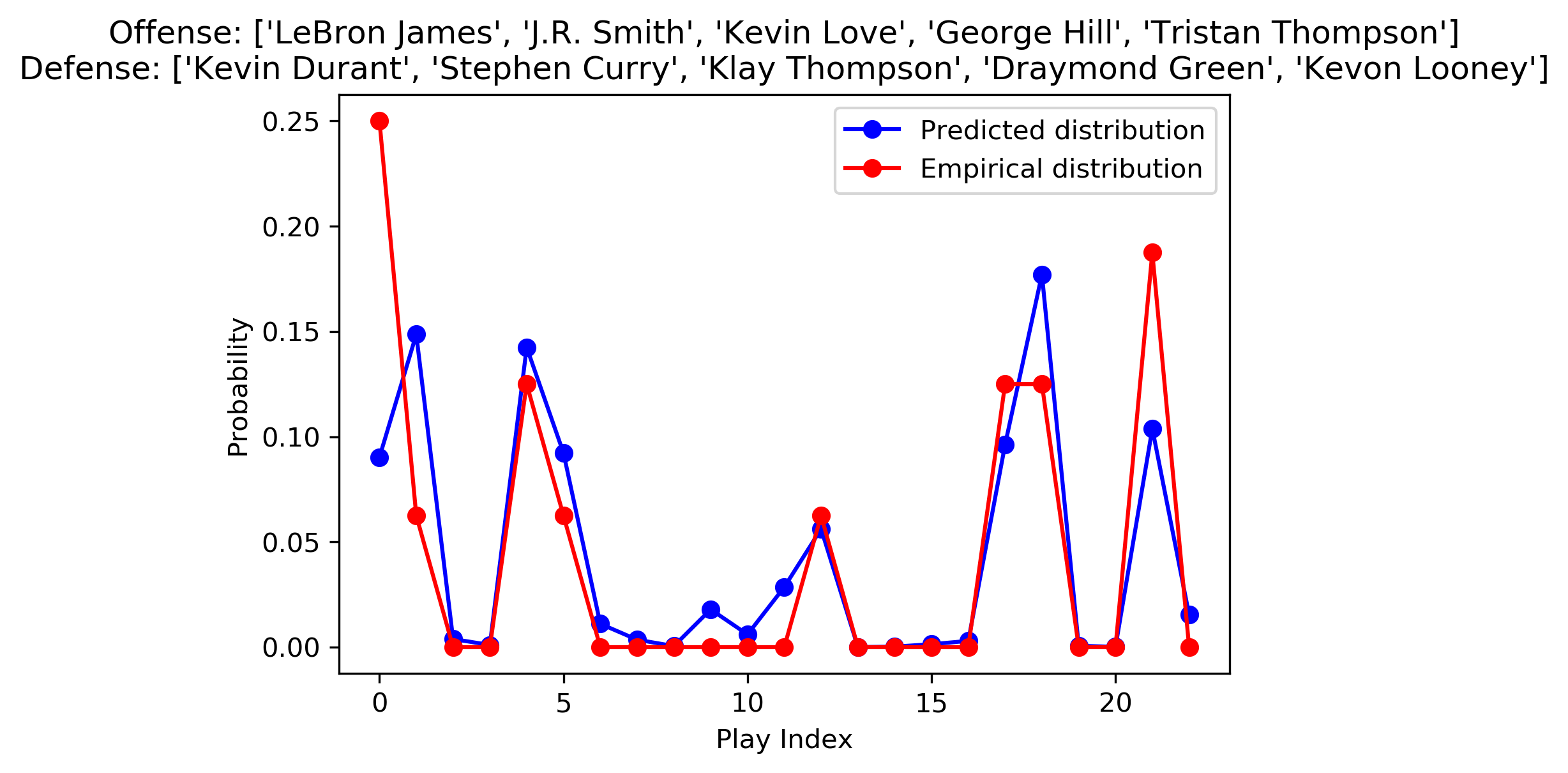}
    \includegraphics[scale=0.425]{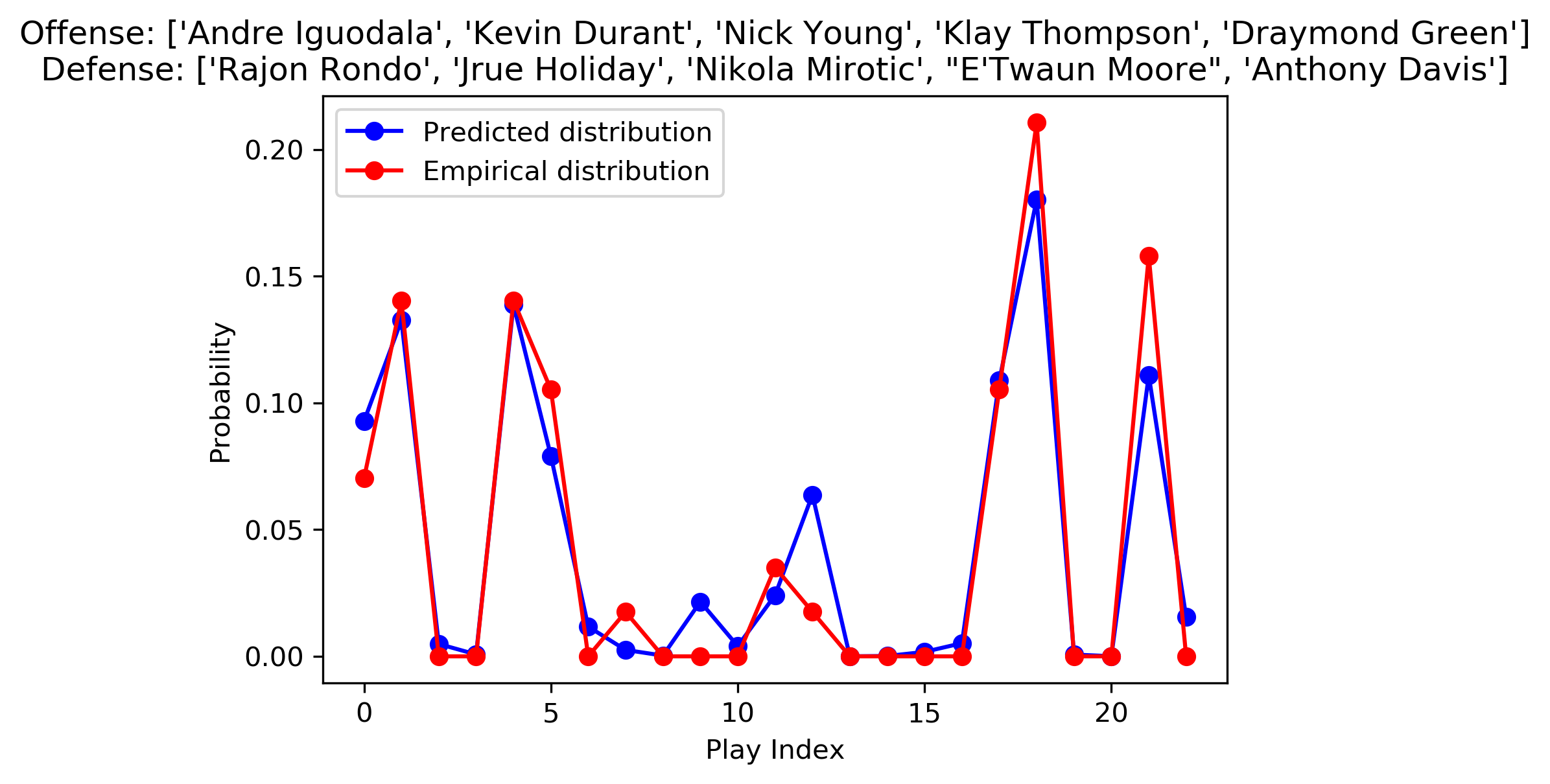}
    \includegraphics[scale=0.425]{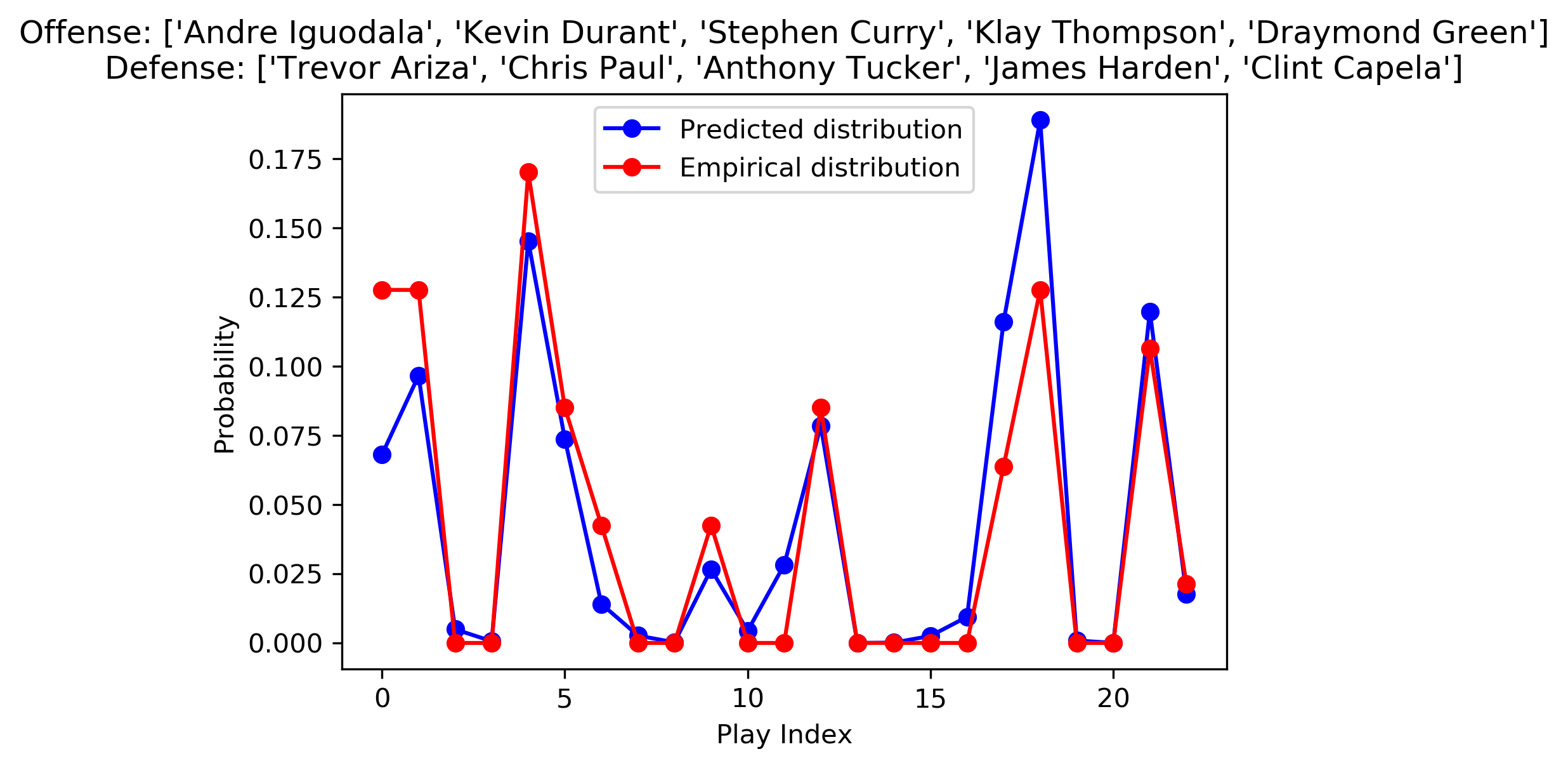}
    \caption{Empirical vs. predicted distributions for various lineup matchups. The outcome corresponding to each play index can be retrieved from Table \ref{tab:2.1}. }
    \label{fig:3.1}
\end{figure*}

\subsection{Exploring Lineup Combinations}
 The results (Table \ref{tab:3.1}) show that even with some crude assumptions, the winner of any given 7-game series and the average margin of victory can be approximated using NBA2Vec embeddings and this neural network model. More accurate game outcomes would require more precise sequence modeling of game-by-game dynamics instead of our current play-by-play treatment; however, this demonstrates the potential of NBA2Vec embeddings and the play outcome predictive network.

\begin{table*} [htp!]
    \centering
    \caption{Results of 1000 simulations of 7-game series of various 2017 playoff matchups vs. ground truth.}
    \begin{tabular}{|c || c c | c | c | c |}
        \hline
         & Team 1 & Team 2 & Series Score & Margin & Team 1 Game Win \% \\
         \hline\hline
        Simulation & Cavaliers & Warriors & 1.69 vs. \textbf{4} & $-8.85$ & 29.8\%\\ 
        Truth & Cavaliers & Warriors & 1 vs. \textbf{4} & $-6.8$ &  20.0\%\\
        \hline 
        Simulation & Jazz & Clippers & \textbf{4} vs. 3.77 & $+0.61$ & 51.5\%\\
        Truth & Jazz & Clippers & \textbf{4} vs. 3 & $+1.14$ & 57.1\%\\
        \hline
        Simulation & Rockets & Thunder & \textbf{4} vs. 2.05 & $+6.63$ & 66.1\%\\
        Truth & Rockets & Thunder & \textbf{4} vs. 1 & $+8.6$ & 80.0\%\\
        \hline 
        Simulation & Warriors & Trailblazers & \textbf{4} vs. 1.06 & $+13.3$ & 79.0\%\\
        Truth & Warriors & Trailblazers & \textbf{4} vs. 0 & $+18$ & 100\%\\
        \hline 
        Simulation & Spurs & Rockets & \textbf{4} vs. 3.96 & $-0.04$ & 50.3\%\\
        Truth & Spurs & Rockets & \textbf{4} vs. 2 & $+5$ & 66.7\%\\
        \hline 
        Simulation & Warriors & Jazz & \textbf{4} vs. 1.51 & $+10.0$ & 72.5\%\\
        Truth & Warriors & Jazz & \textbf{4} vs. 0 & $+15$ & 100\%\\
        \hline 
        Simulation & Wizards & Hawks & \textbf{4} vs. 3.26 & $+1.92$ & 55.1\%\\
        Truth & Wizards & Hawks & \textbf{4} vs. 2 & $+1.17$ & 66.7\%\\
        \hline
        Simulation & Celtics & Wizards & \textbf{4} vs. 3.53 & $+1.11$ & 53.1\%\\
        Truth & Celtics & Wizards & \textbf{4} vs. 3 & $+1$ & 57.1\%\\
        \hline 
        Simulation & Celtics & Cavaliers &3.49 vs. \textbf{4}& $-1.62$ & 46.6\%\\
        Truth & Celtics & Cavaliers & 1 vs. \textbf{4}& $-20$ & 20.0\%\\
        \hline 
    \end{tabular}
    \label{tab:3.1}
\end{table*}

While serviceable as a predictor of game and series outcomes, there is also potential use for NBA2Vec as a lineup optimizer. Given an opposing lineup, this model facilitates selection of a corresponding optimally matched lineup for both offense and defense. This optimization can be accomplished by sampling the model's predicted distribution many times for a given pair of lineups. As an example, we optimize a lineup to face the Golden State Warriors' ``death lineup" (Stephen Curry, Klay Thompson, Andre Iguodala, Kevin Durant, and Draymond Green) for the Houston Rockets, where we fix the first four players (James Harden, Chris Paul, Eric Gordon, Clint Capela) and vary the fifth. From this analysis, we can predict the Rockets' best possible 5th man, and also compare his performance to that of previous starting forward Trevor Ariza. As the simulated win percentages show (Table \ref{tab:3.2}), the ideal 5th man that is currently on the Rockets roster---and is also among the data's 500 most common players---for combating this Warriors lineup is Nene. Compared to Trevor Ariza, only Nene is predicted to add more value. Interestingly, offseason acquisition and superstar Carmelo Anthony is projected to add slightly less value for the Rockets when facing the Warriors than Trevor Ariza (Table \ref{tab:3.2}).   

\begin{table} [htp!]
    \centering
    \caption{Rockets' lineups vs. Warriors for varying ``5th man"}
    \begin{tabular}{|c | c | c |}
        \hline
         Player & Win \% & Avg. Margin of Victory ($\pm 1 \sigma$)\\
         \hline\hline
        Trevor Ariza & 34.4 \% & $-6.69 \pm 17$\\ 
        \hline
        Nene & 37.3\% & $-5.3 \pm 17$\\
        Carmelo Anthony & 33.4\% & $-7.37 \pm 17.1$\\
        Gerald Green & 31.1\% & $-8.47 \pm 17.1$\\
        Joe Johnson & 30.7\% & $-8.1 \pm 16.8$\\
        Brandon Knight & 28.5\% & $-9.22 \pm 16.7$\\
        Michael Carter-Williams & 27.4\% & $-9.58 \pm 16.6$\\
        \hline 
    \end{tabular}
    \label{tab:3.2}
\end{table}


\section{CONCLUSIONS AND FUTURE WORK}

In this study, we have demonstrated that NBA2Vec embeddings can be applied to micro-scale prediction tasks: in particular, predicting the play-by-play outcome distribution given players on the court. Moreover, we have also demonstrated that NBA2Vec also shows potential towards macro-scale prediction tasks, such as in identification of optimal lineups and prediction of game outcomes. In future applications, this could be extrapolated to predictive algorithms for projections of each team’s win-loss record given the players on its roster. 

In addition to applications in predictive tasks, we have shown that the generated NBA2Vec embeddings are able to reveal underlying features of players without using aggregate statistics such as points, FG\%, and assists. Clustering on the embeddings generally groups players in agreement with their position and our priors about their play style/characteristics. Furthermore, the embeddings in part reflect traditional performance metrics, as we are able to show that they correlate at a significant level with box score statistics including rebounds, assists, and field goal rate. Given enough G-League and NCAA training data, player embeddings for potential recruits could also be generated. By examining the nearest neighbor embeddings in the NBA player space, the recruit's ``equivalent NBA player" representation could aid scouts in characterizing him and how he would contribute to a given NBA roster. 

There are various improvements that can be made to potentially extract better player embeddings. Instead of training to predict a singular outcome to every play, a more complex model would train to predict a series of outcomes to each play (e.g. missed shot followed by defensive rebound, shooting foul followed by 2/2 free throws made). To further increase the richness of the embeddings, the network could also be modified to predict the player who commits each action. Finally, with the appropriate player tracking data, a recurrent neural network could be used to take as input time series of player spatial positions and attempt to predict play outcomes and later player spatial positions. Similar to the embeddings generated in this study, these improvements would use only raw data to capture each player's features and ``identity." Ultimately, we envision a future for basketball analytics in which player embeddings allow for unprecedented levels of player characterization, driving predictive models that revolutionize the way front office and coaching decisions are made.  

\section{ACKNOWLEDGEMENTS}
We would like to thank the NBA for organizing the 2018 NBA Hackathon and providing the data for this analysis. We would also like to extend our thanks to Caitlin Chen for her generosity during our trip. 



\bibliography{biblio} 
\bibliographystyle{ieeetr}

\end{document}